\documentclass{optica-article}

\journal{opticajournal} 

\articletype{Research Article}

\def\Fourier{\mathcal{F}}

\begin{document}

\title{Phase-sensitive pump-probe measurement of the complex nonlinear susceptibility of silicon across the direct band edge}

\author{C. D. Cruz,\authormark{1} J. C. Stephenson,\authormark{2} and J. K. Wahlstrand\authormark{1}}

\address{\authormark{1}Nanoscale Device Characterization Division, National Institute of Standards and Technology, Gaithersburg, MD 20899, USA \\ \authormark{2}Associate, Sensor Science Division, National Institute of Standards and Technology, Gaithersburg, MD 20899, USA}

\email{\authormark{*}jared.wahlstrand@nist.gov}

\begin{abstract}
  The nonlinear response of materials, an increasingly important aspect of light-matter interaction, can be challenging to measure in highly absorbing materials.
  Here, we introduce an interferometric technique that enables a direct measurement of the nonlinear complex permittivity in a bulk medium from reflectivity alone.
  We demonstrate the utility of pump-probe supercontinuum (SC) spectral interferometry in reflection by measuring time-dependent variations in the complex dielectric function ($n$, $k$) over the visible wavelength range in bulk silicon.
  Transient phase shifts in the reflected SC due to a near infrared pump pulse allow us to track modifications to $k$; whereas changes in $n$ are derived from transient fluctuations in the reflected SC probe amplitude.
The ultrafast response is attributed to effective two-photon absorption ($\beta$) and Kerr ($n_2$) coefficients.
We observe the onset of strong two-photon absorption as the two-photon energy is tuned through the direct band edge of silicon ($E_1$ = 3.4 eV) for the first time to our knowledge.
This technique allows straightforward spectroscopic measurements of the $\chi^{(3)}$ nonlinear response at the surface of absorbing materials.
\end{abstract}

\section{Introduction}

The optical Kerr effect and two-photon absorption play an increasingly important role in integrated optics \cite{chang_integrated_2022,dutt_nonlinear_2024} and laser processing of materials \cite{phillips_ultrafast_2015,malinauskas_ultrafast_2016}.
The most commonly used techniques for quantifying these aspects of the third-order nonlinear susceptibility $\chi^{(3)}$ measure nonlinear changes in the amplitude, distortion, or direction of an optical beam transmitted through the sample under study \cite{sheik-bahae_high-sensitivity_1989,ferdinandus_beam_2013,negres_two-photon_2002}; however, transmission measurements in highly absorbing bulk materials require preparation of thin samples.
Conventional pump-probe reflection measurements by themselves do not allow access to the change in the full complex dielectric function (they can in combination with transmission measurements, e.g. \cite{downer_ultrafast_1986,kaipurath_optically_2016}).
Directly accessing the complex dielectric function in a pump-probe experiment requires ellipsometry \cite{espinoza_transient_2019} or interferometric techniques \cite{geindre_frequency-domain_1994,misawa_femtosecond_1995,fourment_ultrafast_2018,martinez_sub-picosecond_2021,worle_ultrafast_2021,tamming_frequency_2022}.

The complex reflection coefficient at normal incidence for a single interface between vacuum and a material with complex refractive index $\tilde{n} \equiv n + ik$ is
\begin{equation}
r = \frac{\tilde{n} - 1}{\tilde{n} + 1}.
\end{equation}
Small perturbations of the complex refractive index, for example due to the presence of a pump pulse, change the complex reflectivity according to
\begin{equation}
\Delta r \approx \frac{dr}{d\tilde{n}}  \Delta\tilde{n} = \frac{2}{(\tilde{n} + 1)^2} \Delta\tilde{n},
\label{dreflectivity}
\end{equation}
where $\Delta \tilde{n} = \Delta n + i\Delta k$.
A more experimentally accessible quantity is the normalized change in reflectivity,
\begin{equation}
\frac{\Delta r}{r} = \frac{2\Delta \tilde{n}}{\tilde{n}^2 - 1}.
\label{droverr}
\end{equation}
Ordinary pump-probe experiments measure pump-induced changes in the power reflectivity $R \equiv |r|^2$, i.e.~$\Delta R/R \approx 2 \mathrm{Re}(\Delta r/r)$.
In dielectrics and many semiconductor materials, $n \gg k$ for photon energies up to well above the band edge, and therefore $\tilde{n}^2 - 1 \approx n^2 - 1$ or $\Delta R/R \approx 4 \Delta n/(n^2-1)$.
A standard pump-probe measurement is thus sensitive only to $\Delta n$ in these materials.
A change in the imaginary part of the refractive index $\Delta k$ mostly changes the imaginary part of $\Delta r/r$, which changes the \emph{phase} of reflected light.
The pump-induced phase shift upon reflection is $\Delta \phi = \mathrm{Im}(\Delta r/r) \approx 2\Delta k/(n^2-1)$.

Recently submilliradian pump-probe phase measurements have become possible, enabling transient phase spectroscopy \cite{misawa_femtosecond_1995,tamming_frequency_2022,cruz_pump-probe_2023}.
Here we employ a phase-sensitive pump-probe scheme to measure the wavelength-dependent transient nonlinear response of Si at very short time delay, where the pump pulse overlaps the probe pulse.
Using Eq.~(\ref{droverr}), we extract effective nonlinear coefficients as a function of wavelength and compare to previous results and calculations of the two-photon absorption spectrum in Si.
We use this to measure the two-photon absorption coefficient in bulk Si for two-photon energy (defined as $\hbar \omega_1 + \hbar \omega_2$ for excitation at two frequencies $\omega_1$ and $\omega_2$) well above the direct band edge (3.4 eV), which requires the photon energy of at least one excitation beam to be above the indirect band edge (1.1 eV), at which Si is highly absorbing.
The direct two-photon absorption regime is, with the exception of a few data points around a two-photon energy of 4 eV \cite{reitze_two-photon_1990}, unexplored in Si.
These measurements on Si are of fundamental interest, but we believe the larger impact of this technique will be its ability to measure inhomogeneous flakes and thin films on opaque substrates.

\section{Experiment}

\begin{figure}
    \centering
    \includegraphics[width=12cm]{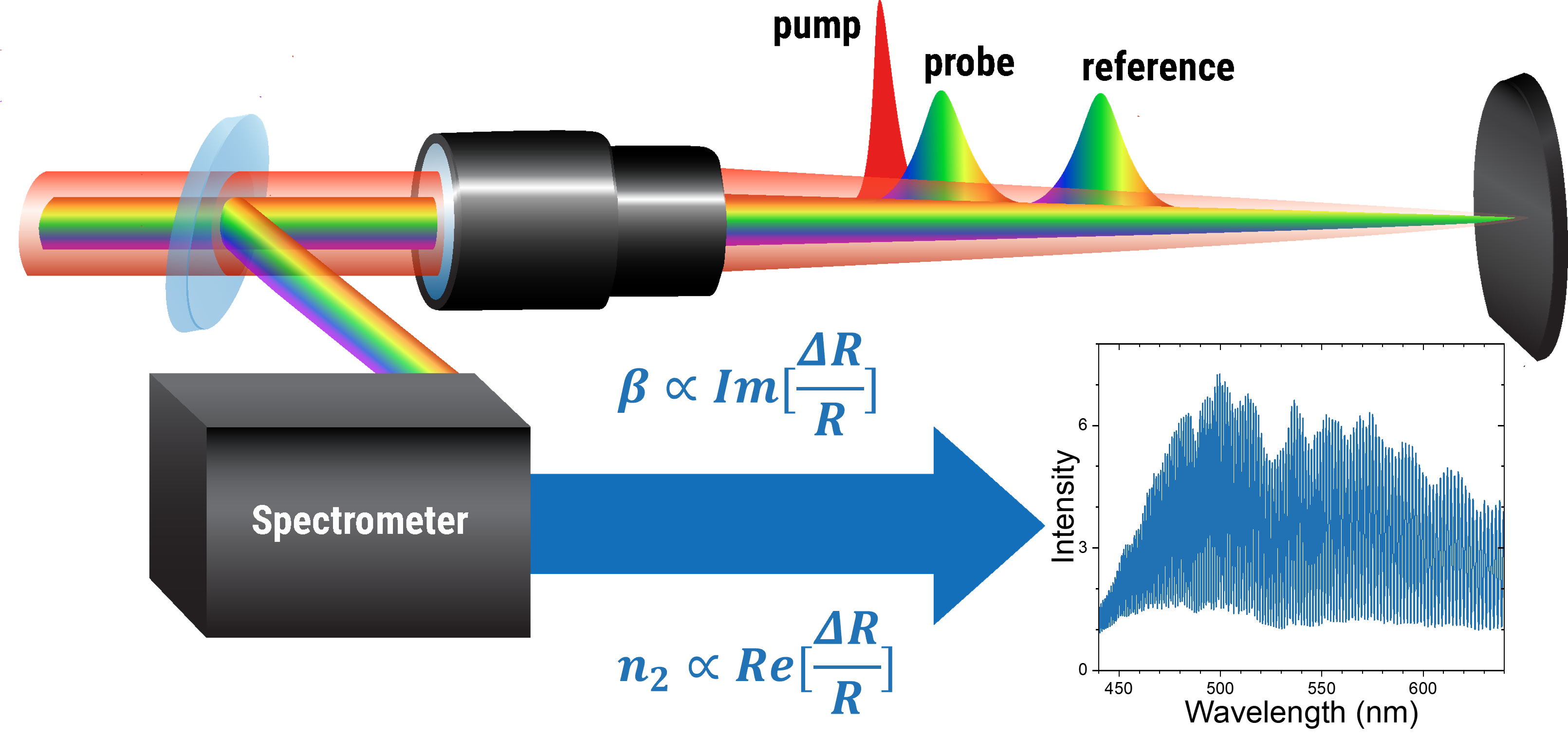}
    \caption{Phase-sensitive pump-probe reflectivity measurements.}
    \label{cartoon}
\end{figure}

The experimental scheme based on pump-probe spectral interferometry is depicted in Fig.~\ref{cartoon}.
We use supercontinuum spectral interferometry \cite{kim_single-shot_2002};
many aspects of the specific experimental apparatus are described in detail in Ref.~\cite{cruz_pump-probe_2023}.
The light source is a Yb:KGW laser (Light Conversion Carbide \cite{NIST_footnote}) that pumps an optical parametric amplifier (OPA, Light Conversion Orpheus \cite{NIST_footnote}).
The laser is operated at 3 kHz repetition rate.
A supercontinuum (SC) probe beam is generated in water by some of the 1.03 $\mu$m output from the laser.
The SC is split into two pulses with a delay $\tau_{pr}$ using a Michelson interferometer.
A wavelength tunable pump beam is generated using the OPA.
The pump and probe beams are combined using a dichroic mirror and focused on the sample with a microscope objective (numerical aperture 0.42).
The central part of the input beam is normally incident on the sample.
Reflected light is sampled with a beamsplitter placed between the objective and the dichroic mirror and is then sent to a spectrometer.
Probe spectra spanning 450-680 nm are collected by a CCD camera at 3 kHz and the pump beam is chopped at 750 Hz.
Captured spectra are sorted and averaged into ``pump on'' and ``pump off'' spectra.
The probe spectra have interference fringes spaced by $1/\tau_{pr}$, and Fourier analysis of these fringes allows extraction of the probe frequency dependent phase and amplitude \cite{takeda_fourier-transform_1982}.

The achievable delay between probe pulses $\tau_{pr}$ has an upper limit set by the resolution of the spectrometer.
In a pump-probe experiment this sets the upper limit on the temporal measurement window, because spectral interferometry requires a reference pulse with an unperturbed spectral phase.
While there are techniques to avoid this limitation for long delay phase spectroscopy \cite{misawa_femtosecond_1995,cruz_pump-probe_2023}, here we are exclusively concerned with the short time response of the sample, within a few hundred femtoseconds of the pump pulse, so we used the maximum time delay possible for our spectrometer resolution, approximately 700 fs.

\begin{figure}
    \centering
    \includegraphics[width=12cm]{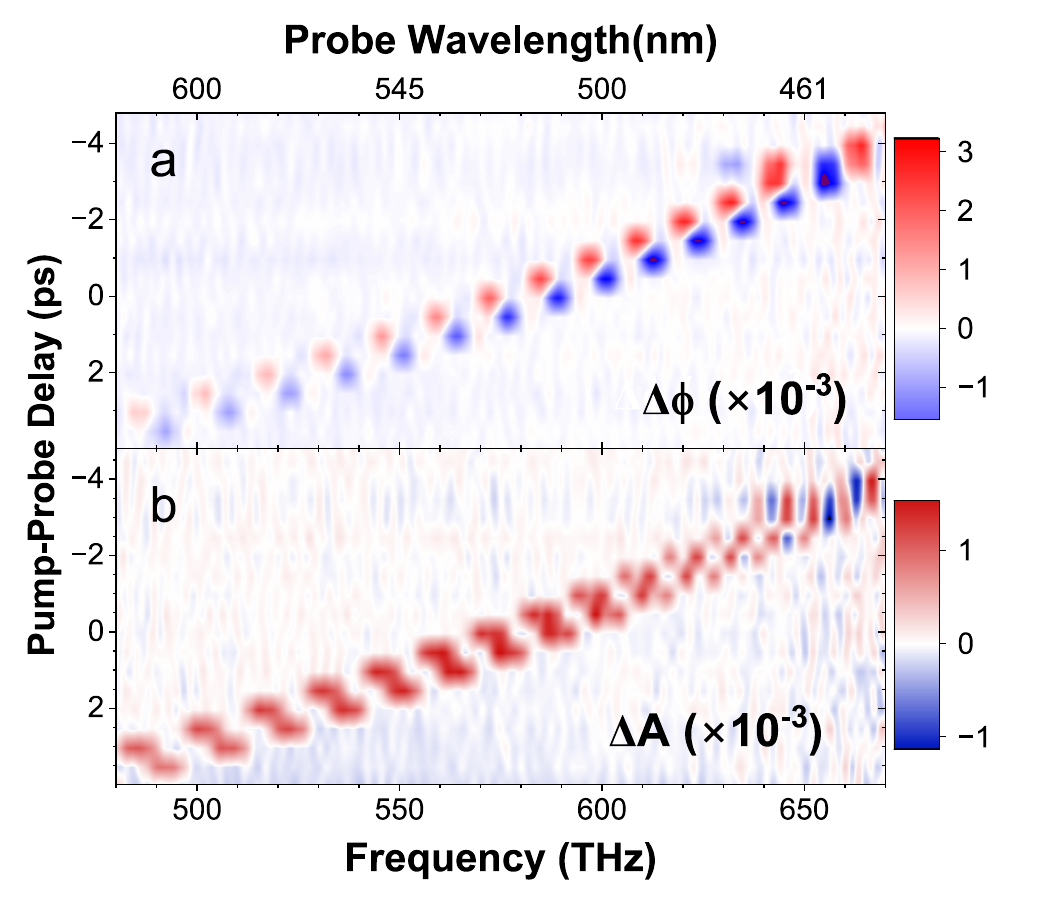}
    \caption{Phase-sensitive pump-probe reflectivity spectrum in bulk Si as a function of pump-probe time delay (with respect to the 570 THz component of the probe pulse) for a 1190 nm pump pulse at 9 mJ/cm$^2$ incident fluence. (a) Extracted pump-induced phase change $\Delta \phi(\omega_{p}, \tau_{ep})$. (b) Extracted pump-induced amplitude change $\Delta A(\omega_p, \tau_{ep})$.}
    \label{raw_traces}
\end{figure}

The supercontinuum pulses are heavily chirped by propagation through the water cell and subsequent optics, with low frequencies arriving first.
For a given pump-probe time delay, the approximately 200 fs long pump pulse therefore overlaps with a small range of probe frequencies.
The extracted phase and amplitude as a function of pump-probe delay $\tau_{ep}$ and probe optical frequency $\omega_p$ are shown in Fig.~\ref{raw_traces} for a (100) oriented Si wafer with parallel pump and probe polarization along the (011) crystal direction.
A pump central wavelength of 1190 nm was used.
Figure \ref{raw_traces}a shows the pump-induced phase shift $\Delta \phi(\omega_{p}, \tau_{ep}) = \mathrm{Im}[\Delta r(\omega_p, \tau_{ep})/r]$, where $\omega_p/(2\pi)$ is the probe frequency and $\tau_{ep}$ is the time delay between the pump pulse and the 570 THz (525 nm) component of the probe pulse.
Figure \ref{raw_traces}b shows the pump-induced normalized field amplitude change $\Delta A(\omega_p, \tau_{ep}) = \mathrm{Re}[\Delta r(\omega_p, \tau_{ep})/r]$ (note that $\Delta A$ is a normalized amplitude change and not a change in absorbance).
The pump-probe delay $\tau_{ep}$ is varied over 10 ps in steps of 500 fs.
To calibrate the pump and probe beam fluence and optimize spatial overlap, we image the sample plane using a second CCD camera.
The width (full width at half maximum) of the probe (pump) beam at the sample surface is 3 $\mu$m (14 $\mu$m).

Transient changes in the phase and amplitude are observed at a probe optical frequency that depends on $\tau_{ep}$ due to the probe chirp.
At a given pump-probe time delay, the closely spaced probe and reference pulses ($\sim 700$ fs) are both affected by the pump pulse at different wavelengths.
When the pump pulse affects the probe pulse but not the reference pulse, the phase shift induced by the pump has a particular sign.
When the pump pulse affects the reference pulse but not the probe, the sign of the pump-induced phase shift is flipped.
In contrast, the pump's effect on the amplitude does not flip sign depending on which pulse is affected.

In spectral interferometry, reliable extraction of the frequency-dependent phase shift of a pulse requires that the reference pulse be unperturbed at the frequencies where the pulse of interest is perturbed by the pump pulse.
Often it is possible to arrange for the reference pulse to not be affected at all by the pump pulse.
Our scheme, where both pulses are affected by the pump pulse, is analogous to (spatial) folded wavefront interferometry, where a beam that propagates through a locally perturbed medium is split and interfered with itself so that an unperturbed region acts as a reference for a phase measurement of the perturbed region (e.g. \cite{chen_direct_2010}).
A detailed description of the time window limitations imposed by the perturbation of the reference pulse can be found in Supplement 1.
The end result of the analysis is that the effect of the pump pulse on each of the probe pulses can be well separated within a $\pm 250$ fs time window.

In both phase and amplitude we observe transient spikes that correspond to the pump pulse overlapping with the probe and reference pulses.
To analyze the data, we draw on previous single-shot supercontinuum spectral interferometry (SSSI) based measurements in gases \cite{kim_single-shot_2002,wahlstrand_absolute_2012}, where knowledge of the probe pulse's spectral phase $\phi_p(\omega)$ was used to calculate the intrapulse-time-dependent phase shift and amplitude change from frequency domain data.
The time domain change in the complex probe envelope is found from the measurements using (see Supplement 1)
\begin{equation}
S(t,\tau_{ep}) = \frac{\Fourier^{-1} [(A_{p}(\omega) + \Delta A(\omega,\tau_{ep})) e^{i[\phi_{p} (\omega) + \Delta \phi(\omega,\tau_{ep})]}]}{\Fourier^{-1} [A_{p} (\omega)e^{i\phi_{p} (\omega)}]} - 1,
\label{textraction}
\end{equation}
where $\Fourier^{-1}$ indicates the inverse Fourier transform, $A_p(\omega)$ is the probe spectral amplitude, $\phi_p$ is the probe spectral phase, $\Delta A(\omega,\tau_{ep})$ is the change in probe spectral amplitude for a certain pump-probe delay $\tau_{ep}$, and $\Delta \phi(\omega,\tau_{ep})$ is the change in probe spectral phase for a certain $\tau_{ep}$.
This is a straightforward generalization of SSSI analysis \cite{kim_single-shot_2002,wahlstrand_optimizing_2016}, which in previous work could assume a pure phase shift in the time domain.
The quantities $\Delta A(\omega,\tau_{ep})$ and $\Delta \phi(\omega,\tau_{ep})$ (shown in Fig.~\ref{raw_traces}) are found from the spectral interferometry analysis described above.
The quantity $A_{p}(\omega)$ is also found directly from the probe interferograms captured by the CCD camera.
We find the last quantity, the probe spectral phase $\phi_p(\omega)$, in the usual way by tracking the frequency of the maximum transient phase shift as a function of time delay $\tau_{ep}$, which is related to $\phi_p(\omega)$ via $\tau_{ep} = \phi_p'(\omega)$ \cite{kim_single-shot_2002,wahlstrand_optimizing_2016}.
The probe spectral phase is well fit by a polynomial $\phi_p(\omega) = \phi_2 (\omega-\omega_0)^2 + \phi_3 (\omega - \omega_0)^3$.
The time resolution of SSSI is in principle given by the inverse of the probe spectral bandwidth, which in our case corresponds to $<10$ fs.
Achieving that ultimate resolution requires perfect knowledge of the probe spectral phase, but we estimate that our time resolution is better than 40 fs.

\section{Results and discussion}

\begin{figure}
    \centering
    \includegraphics[width=12 cm]{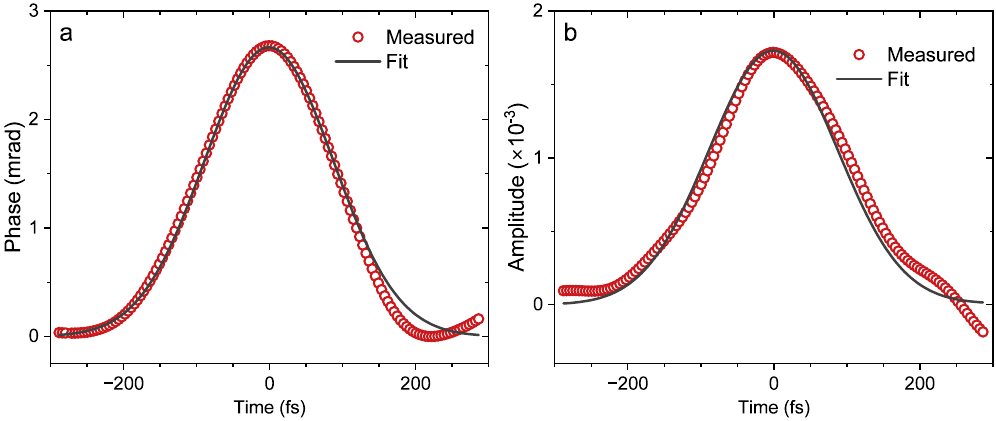}
    \caption{Time domain changes found from data (points) and fits (lines) to Eq.~(\ref{model}) in bulk Si. (a) Time-domain phase change $\Delta \phi(t)$ and (b) Time-domain normalized amplitude change $\Delta A(t)$ pumped at 1190 nm and probed at 492 nm.}
    \label{time_slices}
\end{figure}

For each pump-probe time delay $\tau_{ep}$ (i.e. each row in the subplots shown in Fig.~\ref{raw_traces}), we use Eq.~(\ref{textraction}) to translate the frequency domain phase and amplitude changes to the intrapulse time dependent complex probe envelope change $S(t,\tau_{ep}) = \Delta A(t,\tau_{ep}) + i \Delta \phi(t,\tau_{ep})$.
We truncate the data to include the response of the probe pulse, which results in a $\sim \pm 250$ fs measurement window in $t$ centered on the time at which the pump pulse overlaps the probe pulse.
Examples of extracted time dependent data are shown in Fig.~\ref{time_slices}.
Note that the time axis in Fig.~\ref{time_slices} is the delay within the probe pulse $t$ with respect to the center of the pump pulse, not the pump-probe delay $\tau_{ep}$, the vertical axis in Fig.~\ref{raw_traces}.

Because of the short measurement window, the experiment only captures the transient spike when the pump and probe pulses overlap in time and, potentially, a cumulative signal directly following the pump pulse caused by excitation of free carriers.
When we tune the pump wavelength to photon energy above the indirect band edge of $\approx$1.1 eV, we observe a long-lived amplitude response consistent with previous reflectivity measurements in Si \cite{gunnella_ultrafast_2016,dicicco_broadband_2020}.
Data with pump wavelength 750 nm is shown and briefly discussed in Supplement 1.
Here we intentionally use 1190 nm data to exclude carrier dynamics, which simplifies the analysis.
The shape of the temporal response depends on the pump and probe wavelengths, but in all cases the response at each $\tau_{ep}$ is reasonably well fit by the expression
\begin{equation}
S(t) = \Delta A(t) + i\Delta \phi(t) = S_T\exp(-t^2/\tau_T^2).
\label{model}
\end{equation}
Fits are shown as black solid lines in Fig.~\ref{time_slices}.
In the following text we refer to the term proportional to $S_T$ as the transient response, which is proportional to the approximately Gaussian irradiance profile $I_e(t) = I_0 e^{-t^2/\tau_T^2}$ of our pump pulse.
The time duration parameter $\tau_T$ extracted from the fits is approximately 120 fs (the full width at half maximum is 200 fs), which agrees with the pulse duration we measure from the temporal dependence of the phase shift in transmission in a thin fused silica sample.

\begin{figure}
    \centering
    \includegraphics[width = 12cm]{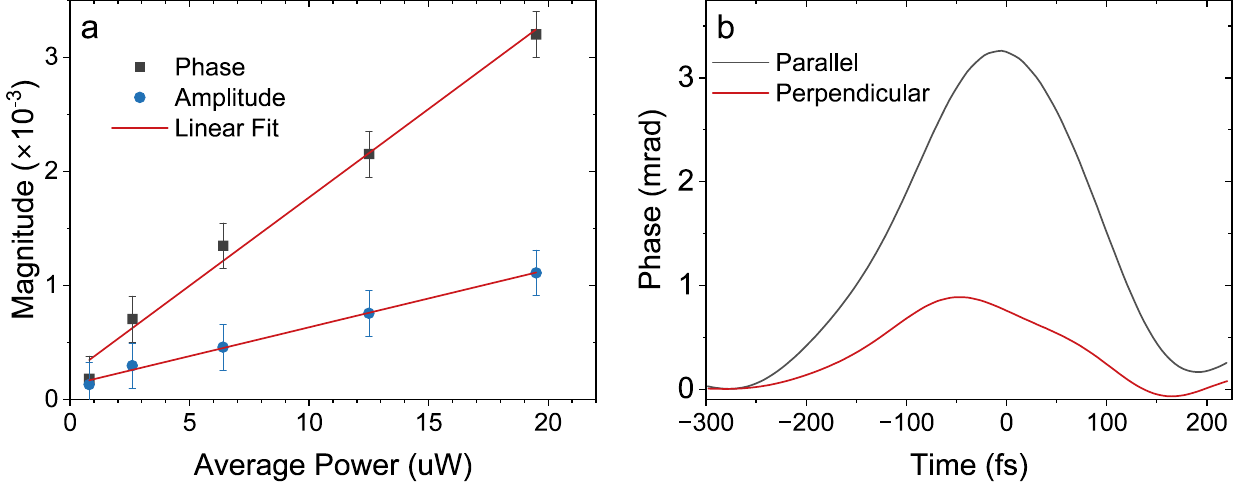}
    \caption{(a) Pump power dependence of the transient phase shift and amplitude change in Si. The error bars are derived from the measurement noise. (b) Polarization dependent $\Delta \phi(t)$. For both plots, the pump wavelength is 1190~nm and probe wavelength is 457~nm.}
    \label{powerpol}
\end{figure}

A transient response of the probe amplitude ($\Delta n$ in reflection when $n \gg k$) that follows the pump irradiance envelope can be modeled by an effective Kerr response \cite{boyd_nonlinear_2008},
\begin{equation}
\Delta n(t) = 2 n_2 I_e(t),
\label{deltan}
\end{equation}
where $n_2$ is the Kerr coefficient for one beam acting on itself.
Note that this expression includes a factor of 2 that is caused by ``weak wave retardation'' \cite{chiao_stimulated_1966,van_stryland_weak-wave_1982,wahlstrand_effect_2013}.
Similarly, a change in the probe phase ($\Delta k$ in reflection when $n \gg k$) can be modeled by a two-photon absorption coefficient,
\begin{equation}
\Delta k(t) = \frac{\lambda}{4\pi} \Delta \alpha (t) = \frac{\lambda}{4\pi} \left[2 \beta I_e(t) \right],
\label{deltak}
\end{equation}
where $\beta$ is the two-photon absorption coefficient and $\lambda$ is the probe wavelength.
We assume that the coefficients $n_2$ and $\beta$ vary slowly enough with wavelength that their values are approximately constant within the approximately 6 nm wide probe bandwidth corresponding to the 200 fs interaction time of the pump pulse and the chirped probe pulse.
The dependence of $S_T$ (phase and amplitude) on pump power (proportional to fluence) is shown in Fig.~\ref{powerpol}a for 1190~nm pump wavelength and is found to be linear within the measurement uncertainty, which is consistent with the expressions above.

We also compared the response with pump and probe polarized parallel and perpendicular to one another.
Time-dependent phase traces are shown in Fig.~\ref{powerpol}b for 1190~nm pump wavelength.
The transient change in both phase and amplitude (latter not shown) is smaller for perpendicular polarization.
In semiconductors, the primary source of an optical nonlinearity that follows the pump irradiance envelope is the bound electronic nonlinearity (a true $\chi^{(3)}$), which is responsible for the optical Kerr effect and two-photon absorption \cite{hutchings_theory_1994}.
The approximate factor of 3 change with polarization is consistent with expectations for the third-order susceptibility for a cubic crystal \cite{hutchings_theory_1994}.

The pump irradiance falls as $I(z) = I_0 e^{-\alpha z}$ within the material because of absorption, and $\Delta n$ and $\Delta k$ drop accordingly.
In an appendix in Ref.~\cite{sabbah_femtosecond_2002}, Sabbah and Riffe show that if the length scale $\delta$ of changes to the optical properties near the surface is much greater than the probe observation depth $d_{\mathrm{obs}} = \lambda/(4\pi n)$, the change of optical constants below the surface can be neglected and the Fresnel formula for reflection, Eq.~(3), can be used.
For the pump and probe wavelengths used, $\delta/d_{\mathrm{obs}} > 500$ putting us firmly within this limit.
Ellipsometry measurements indicated the presence of an approximately $10$ nm thick native oxide layer.
The Kerr coefficient of SiO$_2$ is approximately $3 \times 10^{-16}$ cm$^2$/W \cite{milam_review_1998}, and double passing this thin layer is predicted to produce a peak phase shift $\Delta \phi = 8\pi n_2 I L/\lambda \approx 3\times10^{-5}$ radians, much smaller than the phase shift we observe.

It should be noted that other sources, such as the lattice response, could contribute to a signal that follows a 200 fs pulse intensity profile.
However we expect those contributions to be small, considering the relative size of the coherent phonon signal in Si observed with shorter pulses \cite{sabbah_femtosecond_2002}.
Some caution is also advised concerning the origin of the transient response, because the pump-probe signal when the pulses overlap in time has potential contributions from coherent diffraction effects \cite{smirl_picosecond_1982,wherrett_theory_1983}.
These cause transients in both amplitude and phase \cite{wahlstrand_effect_2013} that depend on details like the pulse chirp.
In the experiments described here, the pump and probe beams are nondegenerate, which massively suppresses these two-beam interference effects because the grating created by the pump and probe beams is time dependent and averages to zero \cite{wherrett_theory_1983,wahlstrand_effect_2013}.
For nondegenerate pulses as used here, the only remaining contribution from two-beam effects is from bound electronic $\chi^{(3)}$, which is unique in that it responds at the optical frequency.
Two-beam effects for a true $\chi^{(3)}$ are responsible for the factor of 2 in Eqs.~(\ref{deltan},\ref{deltak}) \cite{wahlstrand_effect_2013,boyd_nonlinear_2008}.

\section{Extraction of nonlinear coefficients}

\begin{figure}
    \centering
    \includegraphics[width=10cm]{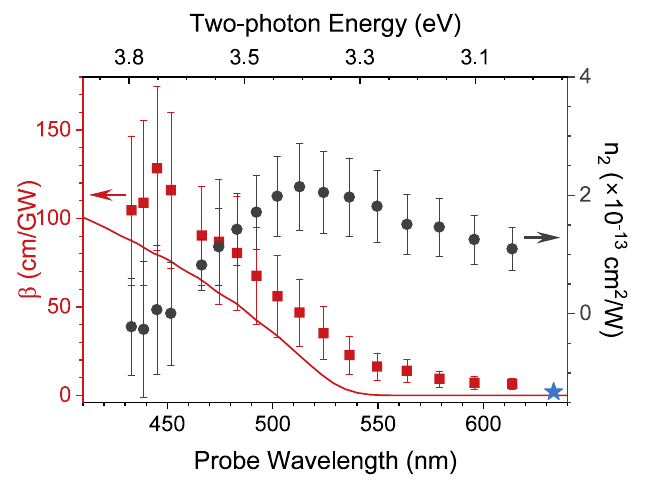}
    \caption{Spectral dependence of measured coefficients (red squares: $\beta$, left axis; black circles: $n_2$, right axis) in Si shown as a function of probe wavelength (bottom axis) and two-photon energy (top axis) at pump wavelengths 1190 nm. The solid line is a calculation of two-photon absorption using a $\mathbf{k} \cdot \mathbf{p}$ model (only including direct electronic transitions). The measured value $\beta$ (degenerate nonlinearity) from Ref.~\cite{bristow_two-photon_2007} is shown as a blue star. }
    \label{spectra}
\end{figure}

We used the previously measured complex refractive index of Si \cite{schinke_uncertainty_2015} to translate from peak phase shift and amplitude change to $\Delta n$ and $\Delta k$ [Eq.~(\ref{droverr})] and used Eqs.~(\ref{deltan},\ref{deltak}) to find effective nonlinear coefficients $n_2$ and $\beta$.
We emphasize that the $n_2$ and $\beta$ coefficients in Eqs.~(\ref{deltan},\ref{deltak}) are nondegenerate coefficients that depend on both pump wavelength and probe wavelength and can be different \cite{fishman_sensitive_2011} from the degenerate coefficients accessed by one-beam experiments \cite{sheik-bahae_high-sensitivity_1989,bristow_two-photon_2007}.
Extracted coefficients are plotted (red points for $\beta$ and black points for $n_2$) as a function of probe wavelength and two-photon energy (sum of pump photon energy and probe photon energy) in Fig.~\ref{spectra}.

\subsection{Uncertainty}

As in most nonlinear optics measurements, the uncertainty in our measurement of the nonlinear coefficients is largely due to uncertainty in the laser irradiance.
In the case of this experiment, we are concerned with the average pump irradiance within the probe spot, which depends on the pulse energy, the spatial beam profile at the sample surface, the temporal profile of the pump pulse, and the overlap between pump and probe spots.
The pump spot at the sample surface was imaged using an InGaAs CCD camera.
A resolution target was used to calibrate the magnification of the image.
The ratio of the summed (dark count subtracted) image to the peak was found and combined with the average power measurement to calculate the peak pump fluence of 9 mJ/cm$^2$.
When setting up the measurement, we peaked up the signal size by adjusting the pump spot position on the sample.
Finally, for the pulse duration we use the parameter $\tau_T = 120$ fs (200 fs full width at half maximum) from the fit of the time domain phase shift.
We estimate the total fractional uncertainty in the peak pump irradiance as 30\%, with inputs from the average power (10\%), the pump spot area (25\%), the pump-probe overlap (10\%), and the pulse duration (10\%).
This contributes to a 30\% fractional uncertainty in each point in Fig.~\ref{spectra}.

In contrast with many other techniques, where the signal-to-noise ratio can be quite high, here the noise in the measurement of $\Delta \phi$ and $\Delta A$ is also a sizable fraction of the uncertainty budget for some data points, particularly at high probe photon energy.
For each data point in Fig.~\ref{spectra} we added (in quadrature) the statistical uncertainty in the peak phase or amplitude from the fit to the 30\% uncertainty due to irradiance.

The absolute uncertainty in each nonlinear coefficient is at least 30\%, but it is important to note that the relative uncertainty between points in the spectra shown in Fig.~\ref{spectra} is smaller because each measurement shares the same pump irradiance.
Thus the uncertainty in pump spot size and pulse duration does not contribute, only the uncertainty from pump-probe overlap (due to probe spatial chirp produced by chromatic aberration) and the phase measurement.
The spectroscopic information in the probe wavelength dependent nonlinear coefficients is therefore of relatively higher quality than the absolute coefficients.
In future work we hope to perform new measurements with an improved estimation of the pump fluence and report coefficients with reduced uncertainty.

\subsection{Comparison to previous work and theory}

For pump wavelength 1190 nm (1.04 eV), the two-photon energy range covers the transition between indirect (phonon-assisted) and direct two-photon absorption (the direct edge is near 3.4 eV).
We find that $\beta$ begins to rise around 3.3 eV, which is similar to the onset of the corresponding one-photon absorption edge \cite{aspnes_dielectric_1983,schinke_uncertainty_2015}.
Our spectral resolution is limited by the pump bandwidth of 10 nm to about 0.01 eV, which is narrower than any features we observe.
We consider the observed two-photon energy dependence to be evidence that we are measuring two-photon absorption, since there is nothing special about the 500 nm probe wavelength in the linear optical properties of Si that would otherwise explain the rapid increase of the transient phase shift with increasing probe photon energy.
Measurements at other pump wavelengths (not shown) also indicated an onset of strong two-photon absorption near 3.3 eV.
We find $\beta \approx 6.5$ cm/GW for two-photon energy around 3.1 eV, which is higher than Z-scan measurements at 3.0 eV two-photon energy by Bristow \emph{et al.} \cite{bristow_two-photon_2007}, shown as a blue star, and other experimental and theoretical work on indirect two-photon absorption in Si \cite{boggess_simultaneous_1986,cheng_full_2011,ziemkiewicz_two-photon_2024}.
Our measurement of $n_2 \approx 1.1\times 10^{-13}$ cm$^2$/W near 3.1 eV is also higher than the Bristow \emph{et al.} value of $4.5 \times 10^{-14}$ cm$^2$/W for their highest reported two-photon energy of 2.4 eV.
The larger values we measure are likely explained by the fact that we measure the nondegenerate susceptibility, which can differ from the degenerate susceptibility \cite{fishman_sensitive_2011} that was measured in Ref.~\cite{bristow_two-photon_2007}.

In direct band gap semiconductors such as GaAs and ZnSe, the spectral dependence of $\beta$ and $n_2$ has been shown to be very consistent with an analytical model based on parabolic valence and conduction bands \cite{christodoulides_nonlinear_2010}.
Unlike GaAs and ZnSe, and even other indirect semiconductors like Ge, the direct band edge in Si at 3.4 eV is not associated with a conduction band minimum at the center of the Brillouin zone ($\Gamma$) aligned with the valence band maximum at $\Gamma$.
Instead the lowest conduction band is concave down in reciprocal space, and the onset of absorption with increasing photon energy occurs at wavevectors well away from the zone center.
We therefore compared the measurements to a calculation of the two-photon absorption spectrum using a 30-band $\mathbf{k} \cdot \mathbf{p}$ model \cite{richard_energy-band_2004,rioux_optical_2012}.
We use the band energy and coupling parameters provided by Richard \emph{et al.} \cite{richard_energy-band_2004} and calculate two-photon absorption spectra using the approach of Rioux and Sipe \cite{rioux_optical_2012}.
The calculation was modified \cite{wahlstrand_unpub} to calculate nondegenerate two-photon absorption spectra, which can result in an enhanced nonlinear response \cite{fishman_sensitive_2011,hannes_higher-order_2019}.
The two-photon absorption spectrum $\beta$ calculated is shown as a solid line in Fig.~\ref{spectra}.
We find very good agreement considering that the theory does not include indirect two-photon absorption.

\section{Conclusion}

In summary, we have demonstrated that the complex nonlinear susceptibility in a highly absorbing bulk medium can be measured through phase-sensitive pump-probe reflectivity spectroscopy.
We hope that our measurement will stimulate theoretical work on direct two-photon absorption in Si.
An important benefit of phase-sensitive pump-probe reflectivity is its ability to measure absolute nonlinearities at surfaces with a small spot size.
This could be used to characterize small samples such as exfoliated flakes and materials that are spatially inhomogeneous.
We emphasize that the technique is not limited to materials with $n \gg k$; using Eq.~(\ref{droverr}) only requires knowledge of $\tilde{n}$.

The technique we demonstrated here has important advantages for characterizing the nonlinear response.
In the transmission geometry, going from the measured phase shift to the change in refractive index requires knowledge of the effective interaction length within the sample.
In reflectivity all that is needed to go from the measurement to the nonlinear change in complex refractive index is the \emph{linear} complex refractive index, which can be easily found through ellipsometry.
Additionally, our analysis does not rely on any assumptions about the 3D beam profile as it propagates through the sample, only the irradiance profile at a single plane (the sample surface).
In Z-scan, the spatiotemporal irradiance profile can be difficult to characterize and result in systematic error.
Thus phase-sensitive pump-probe reflectivity, particularly if its sensitivity can be improved, may offer advantages for transparent samples as well.

Finally, these results on the nonlinear response of the Si surface will support future work on optical nonlinearity measurements of transparent films on Si substrates.
Such films form the basis of many integrated optics platforms and this technique, combined with precise knowledge of the nonlinear response of the Si surface, could be adapted to characterize the optical nonlinearity of films.

\begin{backmatter}
\bmsection{Funding}
NIST Scientific \& Technical Research Services (STRS)

\bmsection{Acknowledgments}
The authors thank Aaron Katzenmeyer for assistance with ellipsometry measurements, Ethan Swagel for help implementing the 30-band $\mathbf{k} \cdot \mathbf{p}$ model, and Valeria Viteri-Pflucker for early experimental contributions.

\bmsection{Disclosures} The authors declare no conflicts of interest.

\bmsection{Data availability} Data underlying the results presented in this paper are not publicly available at this time but may be obtained from the authors upon reasonable request.

\bmsection{Supplemental document} See the supplemental document for supporting content.

\end{backmatter}


\begin{thebibliography}{10}
\bibitem{NIST_footnote}
Certain commercial equipment, instruments, or materials are identified in this paper to foster
understanding. Such identification does not imply recommendation or endorsement by the National
Institute of Standards and Technology, nor does it imply that the materials or equipment identified are
necessarily the best available for the purpose.
\end{thebibliography}

\end{document}


\maketitle

\section{Extracting both amplitude and phase information in SSSI}
In previous implementations of SSSI (e.g. \cite{kim_single-shot_2002,wahlstrand_absolute_2012,wahlstrand_optimizing_2016}), the goal was to extract a time-domain phase shift from the optical Kerr effect or from laser-induced plasma.
In reflectivity, the time-domain signal has both phase and amplitude contributions, and in Si the phase shift comes mostly from two-photon absorption, while the amplitude comes mostly from the optical Kerr effect.
Thus, to extract the full nonlinearity we need to find both phase and amplitude changes from the spectral interferometry measurements.

To explore this we performed numerical simulations, generalizing the approach of \cite{wahlstrand_optimizing_2016}.
We assume that the pump pulse causes both a time domain phase shift $\Delta \Phi(t)$ and normalized amplitude change $\Delta A(t)$.
The probe field in the presence of the pump pulse is
\begin{equation}
E_p(t) = \bar{E}_p(t) \left[\Delta A(t) + 1 \right] e^{i\Delta \Phi(t)},
\label{probefield_time}
\end{equation}
where $\bar{E}_p(t)$ is the unperturbed probe field.
To simplify the following discussion we assume that the effect of the pump pulse is relatively weak, which is well satisfied in the experiment.
We assume that $|\Delta A(t)| \ll 1$ and $|\Delta \Phi(t)| \ll 1$, so that Eq.~(\ref{probefield_time}) becomes
\begin{equation}
  E_p(t) \approx \bar{E}_p(t) + (\Delta A(t) + i \Delta \Phi(t)) \bar{E}_p(t).
  \label{probefield_time_simpler}
\end{equation}

We define the Fourier transform of a function $f(t)$ as
\begin{equation*}
\Fourier[f(t)] \equiv f(\omega) = \int_{-\infty}^{\infty} f(t) e^{i\omega t}dt,
\end{equation*}
and the inverse Fourier transform
\begin{equation*}
\Fourier^{-1} [f(\omega)] \equiv f(t) = \frac{1}{2\pi}\int_{-\infty}^{\infty} f(\omega) e^{-i\omega t}d\omega.
\end{equation*}
In the following, the time domain and frequency domain versions of a quantity will always be clear from the argument.

Taking the inverse Fourier transform of (\ref{probefield_time_simpler}) yields
\begin{equation}
E_p(\omega) = \bar{E}_p(\omega) + \frac{1}{2\pi} \int_{-\infty}^{\infty} \left(\Delta A(\omega-\omega') + i\Delta \phi(\omega-\omega') \right) \bar{E}_p(\omega') d\omega',
\label{probefield_freq}
\end{equation}
where $\Delta A(\omega) = \Fourier \left[\Delta A(t) \right]$, $\Delta \phi(\omega) = \Fourier \left[\Delta \phi(t) \right]$.
The unperturbed frequency domain probe field can be expressed as $\bar{E}_p(\omega) = A(\omega) e^{i\phi_p(\omega)}$, where $A_p(\omega)$ and $\phi_p(\omega)$ are real.

\begin{figure}
    \centering
    \includegraphics[width=12cm]{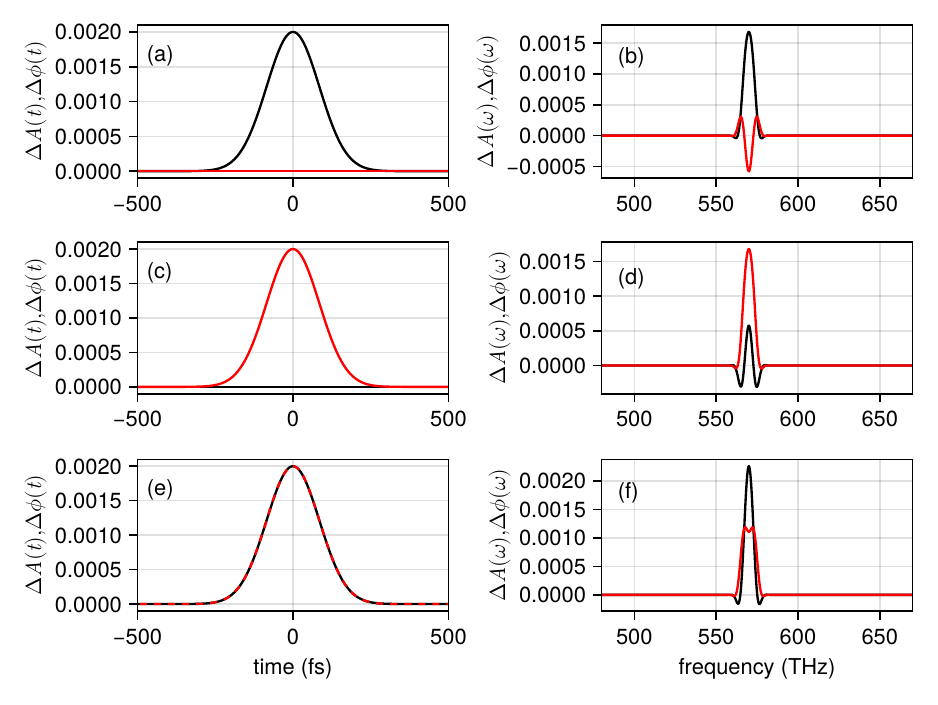}
    \caption{Calculations of phase shifts (black) and amplitude changes (red) for: (a) a pure time domain phase shift and (b) corresponding frequency domain phase and amplitude change; (c) a pure time domain amplitude change and (d) corresponding frequency domain phase and amplitude change; (e) both phase and amplitude changes and (f) corresponding frequency domain phase and amplitude changes. This simulation corresponds to the central row in Fig.~2a (black curves) and Fig.~2b (red curves) in the main text.}
    \label{numerics}
\end{figure}

In \cite{wahlstrand_optimizing_2016}, an analytical expression was derived assuming a Gaussian pump pulse, a nonlinearity consisting only of a phase shift that follows the pump intensity envelope (in transmission, this corresponds to an instantaneous Kerr effect $\Delta n = 2 n_2 I$), and a polynomial probe spectral phase function.
Here we perform numerical simulations using Eq.~(\ref{probefield_freq}) to demonstrate the effect of transient phase shifts and amplitude changes on the frequency domain response.
Calculations of the frequency domain phase shift and amplitude change are shown in Fig.~\ref{numerics}.
We mostly used parameters from the experiment.
For the spectral phase of the probe pulse, we used $\phi_p(\omega) = \phi_2 (\omega-\omega_0)^2 + \phi_3 (\omega-\omega_0)^3$, where $\phi_2 = 2800$ fs$^2$, $\phi_3 = 350$ fs$^3$, and $\omega_0/(2\pi) = 570$ THz.
For the pump pulse, we used a 200 fs FWHM Gaussian pulse centered at $t = 0$ and assumed that its effect on the probe pulse is a phase shift and/or amplitude change that follows the pump intensity envelope.
We assumed a peak time-domain phase shift of 2 mrad and a peak time-domain normalized amplitude change of $2\times 10^{-3}$.

Figure \ref{numerics}b shows a simulation for the pure time-domain phase shift depicted in Fig.~\ref{numerics}a.
In \cite{wahlstrand_optimizing_2016}, which simulated a pump pulse duration of 40 fs, the frequency domain response displayed many oscillations (see Fig. 1bcd in \cite{wahlstrand_optimizing_2016}).
Here because of the longer pump pulse (and different probe chirp), the frequency domain response consists primarily of a peak with a small negative response on either side.
However, note that the pure time-domain phase shift results in a sizable frequency domain amplitude response, and the peak frequency-domain phase shift is about 1.6 mrad, smaller than the 2 mrad peak time domain phase shift.
By reconstructing the time domain response from the frequency domain phase \emph{and amplitude} response and the known probe spectral phase \cite{kim_single-shot_2002}, we can extract the true time domain phase shift.

Figure \ref{numerics}d shows a simulation for the pure time-domain amplitude change depicted in Fig.~\ref{numerics}c.
The calculation of the frequency domain response shows a similar mixing of the time domain amplitude change into frequency domain amplitude change and phase shift.
Figure \ref{numerics}f shows a simulation for a time domain response consisting of both phase and amplitude changes (shown in Fig.~\ref{numerics}e.
The frequency domain response is just the sum of the response in parts b and d of Fig.~\ref{numerics}.
In the experimental data, the relative size of the phase and amplitude response depends on the probe wavelength due to the dispersion of the nonlinear coefficients in Si.

\section{Frequency domain signal when pump pulse affects both probe and reference pulses}
In previous implementations of SSSI (e.g. \cite{kim_single-shot_2002,wahlstrand_absolute_2012,wahlstrand_optimizing_2016}), it was arranged that the pump pulse temporally overlapped only the probe pulse and not the reference pulse.
Because of the required probe spectral bandwidth, very large probe chirp, and equipment limitations, it was not possible to achieve that here.
It is therefore impossible to extract time domain phase and amplitude changes over the full probe pulse duration of roughly 8 ps.
However it \emph{is} possible to extract undistorted information within a small temporal window, analogous to widefield spatial folded wavefront interferometry, as for example used to image underdense plasmas \cite{chen_direct_2010}.
Here we describe the conditions that determine the size of this time window.

\begin{figure}
    \centering
    \includegraphics[width=12cm]{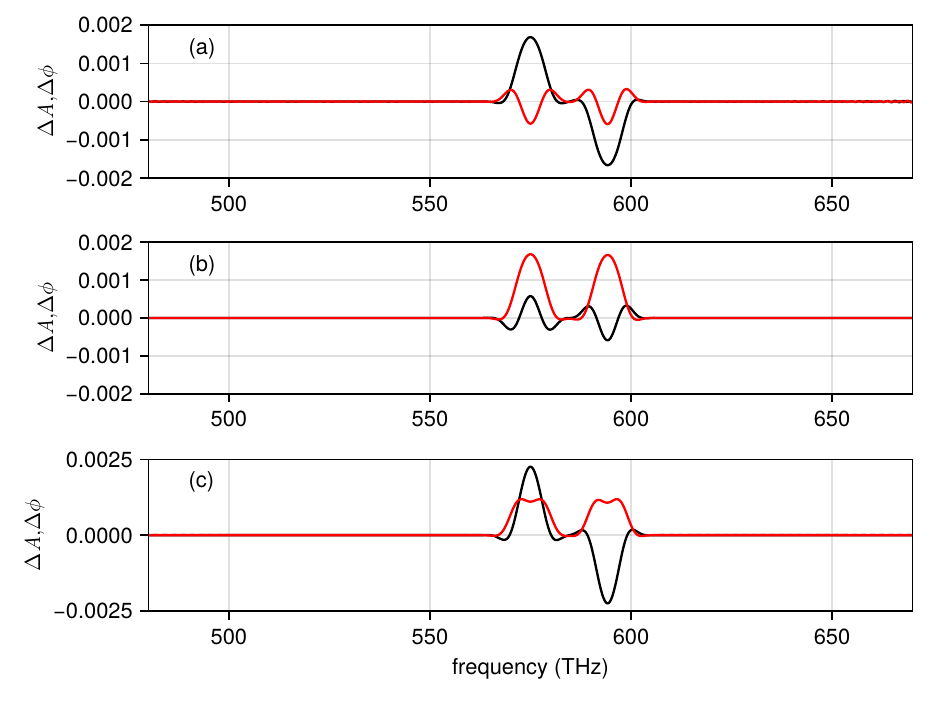}
    \caption{Calculated frequency domain phase shift (black) and amplitude change (red) for: (a) a pure time domain phase shift; (b) a pure time domain amplitude change; (c) both phase and amplitude changes.}
    \label{numerics2}
\end{figure}

Figure \ref{numerics2} shows calculations of the frequency domain phase shift and amplitude change including the effect on both the probe and the reference pulses.
As in the previous section, the effect on the probe is centered at 570 THz.
The reference precedes the probe by 700 fs (the experimental value) and the pump pulse's effect on it is centered near 590 THz.
Compared with the probe, the phase shift for the reference is reversed in sign, while the amplitude carries the same sign.

The width of the frequency domain perturbation is given by the bandwidth of the pump pulse, which extends over about 10 THz for a 200 fs FWHM Gaussian perturbation as assumed in the previous section.
To extract the frequency domain phase and amplitude changes, we need the reference pulse to be undisturbed over this frequency range.
The group delay of frequency components is given by $\phi_p'(\omega) = 2\phi_2(\omega-\omega_0) + 3 \phi_3(\omega-\omega_0)^2$.
Near $\omega_0$, the time delay corresponding to 10 THz is about 350 fs.
This gives the approximate amount of unperturbed time window required to faithfully reconstruct the 200 fs nonlinear phase and amplitude perturbation.
The experimental delay between probe and reference pulses of about 700 fs provides a sufficient buffer to separate the effect of the pump pulse on the probe and reference pulses within a $\pm 250$ fs window of the center of one pulse.

\section{Time domain extraction with both phase and amplitude changes}

The time-dependent change in the complex probe envelope $S(t,\tau_{ep})$ can be found from frequency domain quantities:
\begin{eqnarray}
S(t,\tau_{ep}) &=& \frac{E_p(t,\tau_{ep}) - \bar{E}_p(t)}{\bar{E}_p(t)} \nonumber \\
&=& \frac{E_p(t,\tau_{ep})}{\bar{E}_p(t)} - 1  \nonumber  \\
&=& \frac{\Fourier^{-1} [E_p(\omega,\tau_{ep})]}{\Fourier^{-1} [\bar{E}_p(\omega)]} - 1 \nonumber \\
&=& \frac{\Fourier^{-1} [(A_p(\omega) + \Delta A(\omega,\tau_{ep})) e^{i[\phi_p (\omega) + \Delta \phi(\omega,\tau_{ep})]}]}{\Fourier^{-1} [A_p (\omega)e^{i\phi_p (\omega)}]} - 1, \label{timedomain}
\end{eqnarray}
which is Eq.~(4).

All the quantities in Eq.~(\ref{timedomain}) are measured in the experiment:
\begin{enumerate}
\item $A_p(\omega)$ is the amplitude of $\bar{E}_p$, so it is the square root of the spectrum extracted from the fringes in the pump-off interferogram \cite{takeda_fourier-transform_1982}.
\item $\phi_p(\omega)$ is the spectral phase of $\bar{E}_p$, so it is the polynomial fit we get from looking at the location of the $\tau_{ep}$-dependent transient in the data \cite{kim_single-shot_2002,wahlstrand_optimizing_2016}.
\item $\Delta A(\omega,\tau_{ep})$ is the frequency domain change in amplitude extracted from the interferogram  \cite{takeda_fourier-transform_1982}.
\item $\Delta \phi(\omega,\tau_{ep})$ is the frequency domain phase change extracted from the interferogram  \cite{takeda_fourier-transform_1982}.
\end{enumerate}

\section{Free carrier response}

\begin{figure}
    \centering
    \includegraphics[width=10cm]{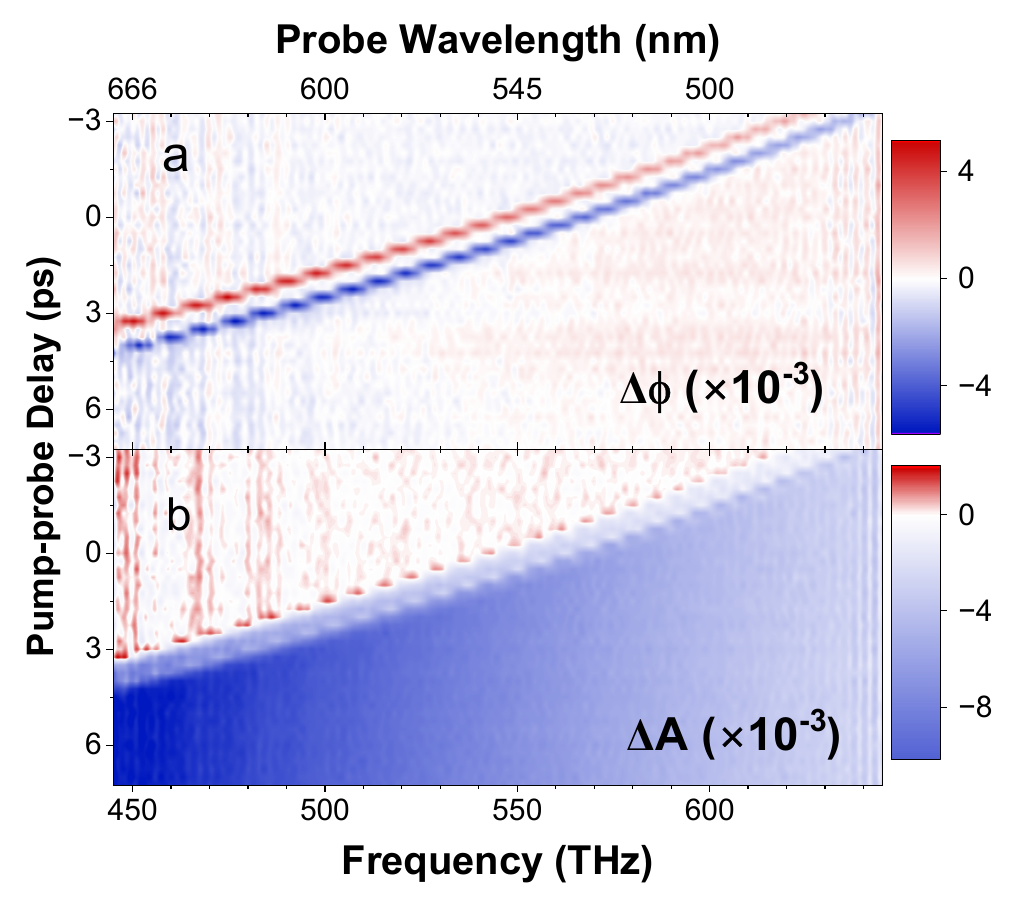}
    \caption{Phase-sensitive pump-probe reflectivity spectrum in bulk Si as a function of pump-probe time delay (with respect to the 570 THz component of the probe pulse) for a 750 nm pump pulse at 50 mJ/cm$^2$ incident fluence. (a) Extracted pump-induced phase change $\Delta \phi(\omega_{p}, \tau_{ep})$; (b) Extracted pump-induced amplitude change $\Delta A(\omega_p, \tau_{ep})$.}
    \label{raw_traces}
\end{figure}

When we tune the pump photon energy to be above the Si band gap, we observe a long-lived free carrier response.
Example data for pump wavelength 750 nm is shown in Fig.~\ref{raw_traces}.
The free carrier response is very strong for 750~nm pump wavelength (Fig.~\ref{raw_traces}), while it is too small to discern for 1190~nm pump wavelength (Fig.~2).
This is explained by the fact that for 750~nm, one-photon absorption is allowed.
The carrier density $N$ at the surface produced by one- and two-photon excitation can be estimated using \cite{sabbah_femtosecond_2002}
\begin{equation}
N \approx (1-R) F \frac{\alpha(\omega_e)}{\hbar \omega_e} + (1-R)^2 F^2 \frac{\beta(\omega_e)}{2 \hbar \omega_e \tau_T},
\label{carrierdensity}
\end{equation}
where $\alpha$ ($\beta$) is the one-photon (two-photon) absorption coefficient, $\omega_e$ is the pump frequency, and $F$ is the pump fluence incident on the sample.
The first term gives the one-photon contribution and the second term the two-photon contribution.

The fluence we used at 750~nm, $F=50$~mJ/cm$^2$, is below the damage threshold of 140~mJ/cm$^2$ for 800~nm, 100~fs pulses reported previously \cite{sjodin_ultrafast_1998}.
Using the known $\alpha(\omega)$ in Si \cite{aspnes_dielectric_1983}, we estimate $N\approx 2 \times 10^{20}$ cm$^{-3}$ from one-photon absorption.
Assuming $\beta = 8$~cm/GW \cite{sabbah_femtosecond_2002}, we calculate $N \approx 1.3\times 10^{20}$ cm$^{-3}$ from two-photon absorption.
In datasets using pump wavelength between 800~nm and 920~nm (not shown) we observed a smaller free carrier response with increasing pump wavelength, as expected from the dispersion of $\alpha$ \cite{aspnes_dielectric_1983} and $\beta$ \cite{bristow_two-photon_2007}.
For 1190~nm pump wavelength (1.04~eV photon energy), two-photon absorption is expected to be the lowest order source of free carriers given the indirect band edge of 1.1~eV.
Using $F = 9$~mJ/cm$^2$ and $\beta = 1.8$~cm/GW \cite{bristow_two-photon_2007} in Eq.~(\ref{carrierdensity}), we find $N = 1.7 \times 10^{18}$~cm$^{-3}$.
The calculated ratio of carrier density for the two wavelengths and fluences, $3.3 \times 10^{20}$~cm$^{-3}$/$1.7 \times 10^{18}$ cm$^{-3}$ $\approx 190$, is in reasonable agreement with the experimentally observed ratio of $>166$.

Our measured amplitude response is consistent with prior pump-probe studies of Si \cite{sjodin_ultrafast_1998,sabbah_femtosecond_2002,gunnella_ultrafast_2016,di_cicco_broadband_2020}.
First, the dependence on probe wavelength is similar:
the long-lived response in amplitude (Fig.~\ref{raw_traces}b) is considerably stronger at long probe wavelengths, consistent with prior pump-probe measurements \cite{gunnella_ultrafast_2016}.
The free carrier response has been found to be well fit by a Drude model (with a correction that accounts for state filling), for which the change in refractive index increases at long wavelengths \cite{sabbah_femtosecond_2002,gunnella_ultrafast_2016}.
Second, the absolute change in reflectivity we observe for 750~nm pump wavelength, $\Delta R/R \approx 1.5 \times 10^{-2}$ near 650~nm probe wavelength for $N \approx 3.3 \times 10^{20}$~cm$^{-3}$, is also reasonably consistent with a set of prior studies that found $\Delta R/R \approx 5.5 \times 10^{-2}$ for $N \approx 1.6 \times 10^{21}$~cm$^{-3}$ \cite{gunnella_ultrafast_2016,di_cicco_broadband_2020}.

Extracting bound electronic nonlinear coefficients $n_2$ and $\beta$ from data that contains a strong free carrier response may be possible, but it will require careful consideration of ultrafast carrier dynamics.
